\documentclass[superscriptaddress,showpacs,nofootinbib,tightenlines,notitlepage]{revtex4-1}
\usepackage{amsmath,verbatim,latexsym,amssymb,graphicx,indentfirst,mathrsfs,mathtools,amsthm,bbm,bm}
\usepackage{hyperref}

\usepackage[multiple]{footmisc}

\usepackage{verbatim,indentfirst}
\usepackage[title,titletoc]{appendix}

\newtheoremstyle{indented}{5pt}{3pt}{\addtolength{\leftskip}{3.5em}}{}{\bfseries}{.}{.5em}{}
\theoremstyle{indented}

\usepackage{soul}

\usepackage{color}
\usepackage[dvipsnames]{xcolor}
\usepackage[normalem]{ulem}

\def\Wext{W_{\rm yield}^{\epsilon}}
\def\Wform{W_{\rm cost}^{\epsilon}}

\newcommand{\Tr}{{\rm Tr}}
\def\id{\mathbbm{1}}
\newcommand{\kB}{k_\mathrm{B}}
\newcommand{\fwd}{\mathrm{fwd}}
\newcommand{\rev}{\mathrm{rev}}

\newcommand*{\ket}[1]{\lvert #1 \rangle}

\newcommand*{\ketbra}[2]{\lvert #1 \rangle\!\langle #2 \rvert}

\begin{document}
\title{Toward physical realizations of thermodynamic resource theories\footnote{
To appear in the conference proceedings for ``Information and Interaction: Eddington, Wheeler, and the Limits of Knowledge,'' in the FQXi subset of Springer's Frontiers book collection.}
}
\author{Nicole~Yunger~Halpern\footnote{E-mail: nicoleyh@caltech.edu}}
\affiliation{Institute for Quantum Information and Matter, Caltech, Pasadena, CA 91125, USA}
\date{\today}

%
%
%

%
%
%
%

%
%
%
%
\begin{abstract}

Conventional statistical mechanics describes large systems and averages over many particles or over many trials. But work, heat, and entropy impact the small scales that experimentalists can increasingly control, e.g., in single-molecule experiments. The statistical mechanics of small scales has been quantified with two toolkits developed in quantum information theory: 
\emph{resource theories} and \emph{one-shot information theory}. 
The field has boomed recently, but the theorems amassed have hardly impacted experiments. Can thermodynamic resource theories be realized experimentally? Via what steps can we shift the theory toward physical realizations? Should we care? I present eleven opportunities in physically realizing thermodynamic resource theories.

\end{abstract}

{\let\newpage\relax\maketitle}
\tableofcontents

%
%
%
%
\section{Introduction}

``This is your arch-nemesis.''

The thank-you slide of my presentation remained onscreen, and the question-and-answer session had begun. I was presenting a seminar about \emph{thermodynamic resource theories} (TRTs), models developed by quantum-information theorists for small-scale exchanges of heat and work~\cite{BrandaoHORS13,FundLimits2}. The audience consisted of condensed-matter physicists who studied graphene and photonic crystals. I was beginning to regret my topic's abstractness.

The question-asker pointed at a listener.

``This is an experimentalist,'' he continued, ``your arch-nemesis. What implications does your theory have for his lab? Does it have any? Why should he care?''

I could have answered better. I apologized that quantum-information theorists, reared on the rarefied air of Dirac bras and kets, had developed TRTs. I recalled the baby steps with which science sometimes migrates from theory to experiment. I could have advocated for bounding, with idealizations, efficiencies achievable in labs. I should have invoked the connections being developed with fluctuation relations~\cite{Aberg13,YungerHalpernGDV14,SalekW15}, statistical mechanical theorems that have withstood experimental tests~\cite{AlemanyR10,MossaMFHR09,ManosasMFHR09,BerutPC13,JunGB14,Blickle06,Saira12,An15}.

The crowd looked unconvinced, but I scored one point: the experimentalist was not my arch-nemesis.

``My new friend,'' I corrected the questioner.

His question has burned in my mind for two years. Experiments have inspired, but not guided, TRTs. TRTs have yet to drive experiments. Can we strengthen the connection between TRTs and the natural world? If so, what tools must resource theorists develop to predict outcomes of experiments? If not, are resource theorists doing physics?

I will explore answers to these questions. I will introduce TRTs and their role in \emph{one-shot statistical mechanics}, the analysis of work, heat, and entropies on small scales. I will discuss whether TRTs can be tested and whether physicists should care. 
I will identify eleven opportunities for stepping TRTs closer to experiments. 
Three opportunities concern what we should realize physically 
and how, in principle, we can realize it.
Six adjustments to TRTs could improve TRTs' realism.
Two opportunities, less critical to realizations, can diversify the platforms with which we might realize TRTs.

The discussion's broadness evokes a caveat of Arthur Eddington's. In 1927, Eddington presented Gifford Lectures entitled \emph{The Nature of the Physical World}. Being a physicist, he admitted, ``I have much to fear from the expert philosophical critic''~\cite{Eddington29}.
Specializing in TRTs, I have much to fear from the expert experimental critic. This paper is intended to point out, and to initiate responses to, the lack of physical realizations of TRTs. Some concerns are practical; some, philosophical. I expect and hope that the discussion will continue.

%
%
%
%
\section{Technical introduction}

The resource-theory framework is a tool applied to quantum-information problems. Upon introducing the framework, I will focus on its application to statistical mechanics, on TRTs. A combination of TRTs and \emph{one-shot information theory} describes small-scale statistical mechanics. I will introduce select TRT results; more are overviewed in~\cite{Goold15}.

%
%
\subsection{Resource theories}
If you have lived in California during a drought, or fought over the armchair closest to a fireplace, you understand what resources are. \emph{Resources} are quantities that have value. Quantities have value when they are scarce, when limitations restrict the materials we can access and the operations we can perform. 
\emph{Resource theories} are simple models used to quantify 
the values ascribed to objects and to tasks by an agent who can perform only certain operations~\cite{CoeckeFS14}. We approximate the agent as able to perform these operations, called \emph{free operations}, without incurring any cost.

Which quantities have value depends on context. Different resource theories model different contexts, associated with different classes of free operations. Classical agents, for example, have trouble creating entanglement. We can model them as able to perform only local operations and classical communications (LOCC). LOCC define resource theories for entanglement. The resource theory for pure bipartite entanglement is arguably the first, most famous resource theory~\cite{HorodeckiHHH09}.\footnote{
Conventional thermodynamics, developed during the 1800s, 
is arguably the first, most famous resource theory. 
But thermodynamics was cast in explicitly resource-theoretic terms only recently.
}
That theory has been used to quantify the rate at which an agent can convert copies of a pure quantum state $\ket{\psi}$ into (maximally entangled) Bell pairs. Using a Bell pair and LOCC, one can simulate a quantum channel. Resource states and free operations can simulate operations outside the restricted class. 

Resource theorists study transformations between states, such as the quantum states $\rho$ and $\sigma$ or the probability distributions $P$ and $Q$. Can the agent transform $\rho$, resource theorists ask, into $\sigma$ via free operations? If not, how much ``resourcefulness'' does the transformation cost? How many copies of $\sigma$ can the agent obtain from $\rho^{\otimes n}$? Can $\rho$ transform into a state $\tilde{\sigma}$ that resembles $\sigma$? 
How efficiently can the agent perform information-processing tasks?
What is possible, and what is impossible? 

So successfully has the resource theory for pure bipartite entanglement answered these questions, quantum information scientists have cast other problems as resource theories. Examples include resource theories for asymmetry~\cite{BartlettRS07,marvian_theory_2013,BartlettRST06}, for stabilizer codes in quantum computation~\cite{VeitchMGE14}, for coherence~\cite{WinterY15}, and for randomness~\cite{CoeckeFS14}. I will focus on resource theories for thermodynamics.

%
%
%
%
\subsection{Thermodynamic resource theories}
Given access to a large equilibrated system, an agent ascribes value to out-of-equilibrium systems. Consider a temperature-$T$ heat bath and a hot ($T' \gg T$) gas. By letting the gas discharge heat into the bath, the agent could extract work from equilibration.\footnote{
More precisely, the agent could extract the capacity to perform work. I will omit the extra words for brevity.}
This work could be stored in a battery, could fuel a car, could power a fan, etc. The resource (the out-of-equilibrium gas) and free operations (equilibration) enable the agent to simulate nonfree operations (to power a car). This thermodynamic story has the skeleton of a resource theory.

The most popular thermodynamic resource theory features an agent given access to a heat bath~\cite{BrandaoHORS13}. The bath has some inverse temperature $\beta := \frac{1}{\kB T}$, wherein $\kB$ denotes Boltzmann's constant. To specify a state $R$, one specifies a density operator $\rho$ and a Hamiltonian $H$ defined on the same Hilbert space: $R  =  (\rho, H)$. (Hamiltonians are often assumed to have bounded, discrete spectra.) To avoid confusion between $R$ and $\rho$, I will refer to the former as a \emph{state} and to the latter as a \emph{quantum state}.

The free operations, called \emph{thermal operations}, tend to equilibrate states. Each thermal operation consists of three steps: (i) The agent draws from the bath a Gibbs state $G$ relative to $\beta$ and relative to any Hamiltonian $H_b$:
$G  =  ( e^{- \beta H_b } / Z,  H_b)$, wherein the partition function $Z :=  \Tr( e^{- \beta H_b} )$ normalizes the state. (ii) The agent implements any unitary that commutes with the total Hamiltonian: 
\begin{align}   \label{eq:UComm}
   [U, H_{\rm tot}]  =  0,
   \qquad  {\rm wherein}  \qquad
   H_{\rm tot}  :=  H  +  H_b  =  (H \otimes \id)  +  (\id  \otimes  H_b). 
\end{align}
This commutation represents energy conservation, or the First Law of Thermodynamics.
(iii) The agent discards any subsystem $A$ associated with its own Hamiltonian. Each thermal operation has the form
\begin{align}
   (\rho, H)  \; \mapsto \;
    \bm{\left( } \Tr_A \left( 
      U \left[   \rho \otimes \frac{ e^{ - \beta H_b } }{Z}   \right]   U^\dag
   \right),
   H  +  H_b  -  H_A  \bm{\right) }.
\end{align}

Each thermal operation decreases or preserves the distance between $R$ and an equilibrium state (for certain definitions of ``distance''~\cite{FundLimits2}). 
As free operations tend to equilibrate states, and as equilibrium states are free, nonequilibrium states are resources. From nonequilibrium states, agents can extract work. 
Work is defined in TRTs in terms of batteries, or work-storage systems.
A battery can be modeled with a two-level \emph{work bit} 
$B_E  =  ( \ketbra{E}{E},  W \ketbra{W}{W} )$ 
that occupies an energy eigenstate $\ket{E} \in \{ \ket{0}, \ket{W} \}$. 
By ``How much work can be extracted from $R$?'' thermodynamic resource theorists mean
(in relaxed notation)
``What is the greatest value of $W$ for which some thermal operation transforms
$R + B_0$ into \mbox{$(Any \; state) + B_W$?''}
Answers involve one-shot information theory.

%
%
%
%
\subsection{One-shot statistical mechanics}
\label{section:OneShotIntro}

Statistical mechanics involves heat, work, and entropy. 
Conventional statistical mechanics describes averages over many trials and over many particles. Yet heat, work, and entropy characterize the few-particle scales that experimentalists can increasingly control (e.g.,~\cite{LiphardtDSTB02,Bustamante05,Faucheux95,ChengSHEL12,Pekola15}), as well as single trials. Small scales have been described by TRTs (e.g.,~\cite{FundLimits2,BrandaoHNOW14,GourMNSYH13,YungerHalpernR14,YungerHalpern14,LostaglioJR14,Ng14}). These results fall under the umbrella of \emph{one-shot statistical mechanics}, an application of the \emph{one-shot information theory} that generalizes \emph{Shannon theory}~\cite{RennerThesis}. 

I will introduce Shannon theory and the \emph{Shannon entropy} $H_{\rm S}$ with the thermodynamic protocol of compressing a gas. The quantum counterpart of $H_{\rm S}$ is the \emph{von Neumann entropy} $H_{\rm vN}$. Both entropies, I will explain, quantify efficiencies in the large-scale \emph{asymptotic limit} of information theory, which relates to the \emph{thermodynamic limit} of statistical mechanics. Outside these limits, one-shot entropic quantities quantify efficiencies. I will introduce these quantities, then illustrate with work performance.

\emph{Shannon theory} concerns averages, over many copies of a random variable or over many trials, of the efficiencies with which information-processing tasks can be performed~\cite{CoverT12}. Many average efficiencies are functions of the Shannon entropy $H_{\rm S}$ or the von Neumann entropy $H_{\rm vN}$.
Examples include the performance of work.

Consider a classical gas in a cylinder capped by a piston. 
Suppose that the gas, by exchanging heat through the cylinder with a temperature-$T$ bath, has equilibrated. Imagine compressing the gas quasistatically (infinitely slowly) from a volume $V_i$ to a volume $V_f$. One must perform work against the gas particles that bounce off the piston. The work required per trial averages to the change $\Delta F$ in the gas's \emph{Helmholtz free energy}:
\begin{align}
\label{eq:FE}
   \langle W \rangle  =  \Delta F,
   \qquad  {\rm wherein}  \qquad
   F = \langle E \rangle - T S.
\end{align} 
$\langle E \rangle$ denotes the average, over infinitely many trials, of the gas's internal energy; and $S$ denotes the gas's statistical mechanical entropy~\cite{AndersV15}.


The statistical mechanical entropy of a classical state has been cast in terms of the information-theoretic Shannon entropy~\cite{AndersV15,Jarzynski97PRE}. 
The \emph{Shannon entropy} of a discrete probability distribution $P = \{ p_i \}$ is defined as
\begin{align}
\label{eq:ShannonH}
   H_{\rm S} (P)   :=
   - \sum_i  p_i  \ln ( p_i ).
\end{align}
The Shannon entropy quantifies our ignorance about a random variable $X$ whose possible outcomes $x_i$ are distributed according to $P$. The $P$ relevant to the gas is the distribution over the microstates that the gas might occupy: $P = \{ e^{ - \beta E_i } / Z \}$, wherein $E_i$ denotes the energy that the gas would have in Microstate $i$ and the partition function $Z$ normalizes the state. 
The \emph{surprise} $-\ln( p_i )$ quantifies the information we gain upon discovering that the gas occupies Microstate $i$. The surprise quantifies how much the discovery shocks us.
Imagine learning which microstate the gas occupies in each of $n$ trials. $H_{\rm S} (P)$ equals the average, in the limit as $n \to \infty$, of the per-trial surprise. The statistical mechanical entropy $S  =  \kB  H_{\rm S} (P)$ is proportional to the average of our surprise. 
$H_{\rm S}$ quantifies the average, over $n \to \infty$ trials, of the efficiency with which a gas can be compressed quasistatically. 

Statistical mechanical averages over $n \to \infty$ trials are related to the \emph{thermodynamic limit}, which is related to the \emph{asymptotic limit} of information theory.
The thermodynamic limit is reached as a system's volume and particle number diverge 
($V, N \to \infty$), while $V / N$ remains constant.
Imagine performing the following steps $n$ times:
(i) Prepare a system characterized by a particular volume $V$ and particle number $N$. 
(ii) Measure a statistical mechanical variable, such as the energy $E$.
Let $E_\gamma$ denote the outcome of the $\gamma^{\rm th}$ trial.
Let $\langle E \rangle$ denote the average, over $n \to \infty$ trials, of $E$.
In the thermodynamic limit, the value assumed by $E$ in each trial 
equals the average over trials:
$E_\gamma  \to  \langle E \rangle$~\cite{Schroeder00}.

The thermodynamic limit of statistical mechanics relates to an asymptotic limit of information theory. 
In information theory, the Shannon entropy quantifies averages over $n \to \infty$ copies of a probability distribution $P = \{ p_i \}$. A random variable $X$ can have a probability $p_i$ of assuming value the $x_i$. 
Consider compressing $n$ copies of $X$, jointly, into the fewest possible bits. 
These $X$'s are called \emph{i.i.d.,} or \emph{independent and identically distributed}: 
no variable influences any other, and each variable is distributed according to $P$. 
In the \emph{asymptotic limit} as $n \to \infty$, the number of bits required per copy of $X$ approaches $H_S(P)$. 
Functions of $H_{\rm S}$ quantify averages of the efficiencies with which many classical tasks can be performed in the asymptotic limit~\cite{CoverT12}. 

The asymptotic average efficiencies of many quantum tasks depend on the \emph{von Neumann entropy} $H_{\rm vN}$~\cite{NielsenC10}. Let $\rho$ denote a density operator that represents a system's quantum state. The von Neumann entropy 
\begin{align}
   H_{\rm vN} (\rho)   :=
   - \Tr \bm{(} \rho \log ( \rho ) \bm{)}
\end{align}
has an operational interpretation like the Shannon entropy's:
Consider preparing $n$ copies of $\rho$. Consider projectively measuring each copy relative to the eigenbasis of $\rho$. The von Neumann entropy quantifies the average, in the limit as $n \to \infty$, of our surprise about one measurement's outcome~\cite{NielsenC10}.

$H_{\rm S}$ and $H_{\rm vN}$ describe asymptotic limits, but infinitely many probability distributions and quantum states are never processed in practice.
Nor are infinitely many trials performed. 
Into how few bits or qubits can you compress finitely many copies of $P$ or $\rho$?
Can you compress into fewer by allowing the compressed message to be decoded inaccurately?
How much work must you invest in finitely many gas compressions, e.g., in one ``shot''? 
Such questions have been answered in terms of the \emph{order-$\alpha$ R\'enyi entropies} $H_\alpha$ and \emph{R\'enyi divergences} $D_\alpha$~\cite{Renyi61,vanErvenH12}.

$H_\alpha$, parameterized by $\alpha \in [0, \infty)$, 
generalizes the Shannon and von Neumann entropies. 
In the limit as $\alpha \to 1$, 
$H_\alpha(P) \to H_{\rm S}(P)$ (if the argument is a probability distribution), 
and $H_\alpha(\rho)  \to  H_{\rm vN}(\rho)$ (if the argument is a quantum state).
Apart from $H_1$, two R\'enyi entropies will dominate our discussion:
\begin{align}
   H_\infty  =  - \log ( p_{\rm max} )
\end{align}
depends on the greatest probability 
$p_{\rm max}$ in a distribution $P$ 
or on the greatest eigenvalue $p_{\rm max}$ of a quantum state $\rho$.
$H_\infty$ relates to the work that a TRT agent must invest to create a state 
$(\rho, H)$~\cite{GourMNSYH13,FundLimits2}.
The work extractable from $(\rho,  H)$ relates to 
\begin{align}
   H_0  =  \log (d)
\end{align}
wherein $d$ denotes the support of $P$ (the number of nonzero $p_i$'s) 
or the dimension of the support of $\rho$ (the number of nonzero eigenvalues of $\rho$)~\cite{GourMNSYH13,FundLimits2}.
The R\'enyi divergences $D_\alpha$ generalize the relative entropy $D_1$. 
They quantify the discrepancy between two probability distributions 
or two quantum states~\cite{NielsenC10,vanErvenH12}.

Variations on the $H_\alpha$ and $D_\alpha$ have been defined and applied (e.g.,~\cite{Petz86,RennerThesis,Datta09,Tomamichel12}).
Examples include the \emph{smooth R\'enyi entropies}~\cite{RennerThesis}
$H_\alpha^\epsilon$ 
and the \emph{smooth R\'enyi divergences} $D_\alpha^\epsilon$.
I will not define them, to avoid technicalities.
But I will discuss them in Sec.~\ref{section:Embezzle} and Sec.~\ref{section:Error},
after motivating smoothing with efficiency.

One-shot entropic quantities quantify the efficiencies 
with which ``single shots'' of tasks can be performed. Information-processing examples include data-compression rates~\cite{RennerThesis} and channel capacities~\cite{BuscemiD09}. Quantum-information tasks include quantum key distribution~\cite{RennerThesis}, the distillation of Bell pairs from pure bipartite entangled states $\ket{\psi}$, and the formation of $\ket{\psi}$ from Bell pairs~\cite{BrandaoD09}. Thermodynamic efficiencies include the minimum amount $W_{\rm cost}$ of work required to create a state $R$ and the most work $W_{\rm yield}$ extractable from $R$~\cite{FundLimits2}.

The $W_{\rm cost}$ and $W_{\rm yield}$ of quasiclassical states have been calculated with TRTs. A \emph{quasiclassical} state $R = (\rho, H)$ has a density operator that commutes with its Hamiltonian:\footnote{
Many TRT arguments rely on quasiclassicality as a simplifying assumption. 
After proving properties of quasiclassical systems, thermodynamic resource theorists attempt generalizations to coherent states.}
$[\rho, H] = 0$.
According to~\cite{FundLimits2},
the work extractable from one copy of $R$ is\footnote{
$D_0$ and $D_\infty$ are called $D_{\rm min}$ and $D_{\rm max}$ in~\cite{FundLimits2}. I follow the naming convention in~\cite{GourMNSYH13}: if $P$ denotes a $d$-element probability distribution and $u$ denotes the uniform distribution $( \frac{1}{d},  \ldots,  \frac{1}{d} )$, then
$D_\infty (P ||u) = \log (d) - H_\infty (P )$, and 
$D_0 (P ||u) = \log (d) - H_0 (P )$.}
$W_{\rm yield} (R)  =  \kB T \:  D_0 \left( \rho \; || \; e^{ -\beta H } / Z \right)$. 
Creating one copy of $R$ costs
$W_{\rm cost} (R)  =  \kB T \:   D_\infty \left( \rho \; || \; e^{ -\beta H } / Z \right)$. 

One-shot efficiencies have been generalized to imperfect protocols whose outputs resemble, but do not equal, the desired states~\cite{RennerThesis}. 
Suppose that a TRT agent wants to create $R$. 
The agent might settle for creating any $\tilde{R}  =  (\tilde{\rho},  H)$ whose density operator $\tilde{\rho}$ is close to $\rho$. 
That is, the agent wants for $\tilde{\rho}$ to lie within a distance $\epsilon \in [0, 1]$ of $\rho$: 
$\mathcal{D}( \tilde{\rho}, \rho)  \leq  \epsilon$. 
(Distance measures $\mathcal{D}$ are discussed in Sec.~\ref{section:Embezzle}.)  
$\epsilon$ is called a \emph{smoothing parameter}, \emph{error tolerance}, or \emph{failure probability}.
The least amount of work required to create any $\tilde{R}$ 
depends on the smooth order-$\infty$ R\'enyi divergence $D^\epsilon_\infty$~\cite{Datta09}:
$W^\epsilon_{\rm cost} (R)  =  
\kB T \:   D^\epsilon_\infty \left( \rho \; || \; e^{ -\beta H } / Z \right)$. 
The most work extractable from $R$ with a similar faulty protocol depends on the smooth order-$0$ R\'enyi divergence:
\mbox{$\Wext (R)  =  
\kB T \:   D^\epsilon_0 \left( \rho \; || \; e^{ -\beta H } / Z \right)$}~\cite{FundLimits2}.

Let us compare these one-shot work quantities with conventional thermodynamic work. Compressing a gas in the example above costs an amount 
$ W_{\rm cost}^{\rm th}  =  \Delta F$ of work. 
Consider expanding the gas from $V_f$ to $V_i$. The gas performs an amount $W_{\rm yield}^{\rm th}   =  \Delta F$ of work on the piston. In thermodynamics, the work cost equals the extractable work. In one-shot statistical mechanics, $W_{\rm cost}(R)$ is proportional to $D_\infty$, whereas $W_{\rm yield}(R)$ is proportional to $D_0$. 
From the one-shot results, the thermodynamic results have been recovered~\cite{FundLimits2}.

%
%
%
%
\section{Opportunities in physically realizing thermodynamic resource theories}
\label{section:Challenges}

I have collected eleven opportunities from conversations, papers, and contemplation.
Opportunity one concerns the question ``Which aspects of TRTs merit realization?''
Two and three concern how, in principle, these aspects can be tested.
I call for expansions of the TRT framework, intended to enhance TRTs' realism, 
in the next six sections.
I invite more-adventurous expansions in the final two sections.

%
%
%
%
\subsection{What merits realization? How, in principle, can we realize it?}

\emph{Prima facie}, single-particle experiments exemplify
the one-shot statistical mechanics developed with TRTs.
But many-body systems could facilitate tests.
I explain how in section one.
Section two concerns our inability to realize 
the optimal efficiencies described by TRT theorems.
Section three concerns which steps one should perform in a lab,
and how one should process measurements' outcomes,
to realize TRT results.

%
%
%
%
\subsubsection{What would epitomize realizations of the one-shot statistical mechanics developed with TRTs?}
\label{section:OneShot}

Many TRT results fall under the umbrella of \emph{one-shot statistical mechanics},  said to describe small scales. 
In conventional statistical mechanics, small scales involve few particles and small volumes. 
Yet much of the mathematics of one-shot statistical mechanics can describe many particles and large volumes. 
To realize TRT results physically, we must clarify what needs realizing.
I will first sketch why one-shot statistical mechanics seems to describe single particles. After clarifying in detail, I will demonstrate how one-shot statistical mechanics can describe many-body systems. I will argue that physically realizing one-shot statistical mechanics involves observing discrepancies between R\'enyi entropic quantities. This argument implies that entangled many-body states could facilitate physical realizations of several one-shot results.

At first glance, ``one-shot'' appears to mean ``single-particle'' in TRT contexts. 
Conventional statistical mechanics describes collections of about $10^{24}$ particles. 
Most of such particles behave in accordance with averages. 
(Average behavior justifies our modeling of these collections with statistics, rather than with Newtonian mechanics.)
The Shannon and von Neumann entropies ($H_1$) quantify averages over large numbers. 
Other R\'{e}nyi entropies ($H_{\alpha \neq 1}$) quantify work in one-shot statistical mechanics. Hence one-shot statistical mechanics might appear to describe single particles.

What one-shot statistical mechanics describes follows from what one-shot information theory describes, since the former is an application of the latter. One-shot information theory governs the simultaneous processing of arbitrary numbers $n$ of copies of a probability distribution $P$ or of a quantum state $\rho$. 
Hence one-shot statistical mechanics could describe few particles if one copy of $R = (\rho, H)$ described each particle. 

One copy can, but need not, describe one particle. Qubits exemplify the ``can'' statement. The \emph{qubit}, the quantum analog of the bit, is a unit of quantum information. Qubits form the playground in which many one-shot-statistical-mechanics theorems are illustrated (e.g.,~\cite{YungerHalpernGDV14,LostaglioKJR15}). A qubit can incarnate as the state of a two-level quantum system. For example, consider an electron in a magnetic field $\mathbf{B} = B \hat{z}$. The Zeeman Hamiltonian 
 $H_{\rm Z}   \propto   \sigma_z$ governs the electron's spin, wherein $\sigma_z$ denotes the Pauli $z$-operator. 
 $H_{\rm Z}$ defines two energy levels, and a density operator $\rho$ represents the electron's spin state. 
If $R = (\rho, H_{\rm Z})$,
the one-shot quantities $W_{\rm cost}(R)$ and $W_{\rm yield}(R)$ represent the work cost of creating, and the work extractable from, one electron. 
Hence one copy of a state $R$ can describe one particle; 
and, as argued above, one-shot statistical mechanics can describe single particles.

Yet one copy of $R$ can represent the state of a system of many particles. 
For example, let $R = (\rho, H_{XX})$ denote the state 
of a chain of spin-$\frac{1}{2}$ particles. 
Such a chain is governed by the Hamiltonian
$H_{XX}  =  - \frac{J_{\rm ex} }{2}  
\sum_j   ( S^+_j S^- _{j + 1}  +  S^+_{j + 1}   S^-_j )$, 
wherein $J_{\rm ex}$ denotes the uniform coupling's strength; 
the index $j$ can run over the arbitrarily many lattice sites;
and $S^+_j$ and $S^-_j$ denote the $j^{\rm th}$ site's raising and lowering operators~\cite{Mazza15}.
This many-body state $R$ can obey TRT theorems derived from one-shot information theory. Hence one-shot statistical mechanics can describe many-body systems. 

My point is not that one-shot functions can be evaluated on 
\mbox{$n > 1$} i.i.d. copies of a state,
denoted by \mbox{$R^n  =  ( \rho^{\otimes n},  \oplus_{i = 1}^n H)$}. 
One-shot functions, such as $H_\alpha$ or $W_{\rm cost}$, are evaluated on $R^n$ in arguments about the asymptotic limit.
Asymptotic limits of one-shot functions 
lead to expressions reminiscent of conventional statistical mechanics
[e.g., Eqs.~\eqref{eq:FE}], 
which describes many particles~\cite{FundLimits2,GourMNSYH13,LostaglioJR14}.
Therefore, $R^n$ is known to be able to represent the state of a many-particle system.
But my point is that $R$, not only $R^n$, can represent the state of a many-particle system. 
$W_{\rm cost}( R )$ can equal the work cost of creating ``one shot'' of a many-particle state~\cite{YungerHalpernR14,YungerHalpern14}. 

Therefore, few-particle systems would not necessarily epitomize physical realizations of one-shot statistical mechanics. What would? Quantum states $\rho$ whose one-shot entropic quantities
($D_{\alpha \neq 1}$, 
$H^\epsilon_0$, $H^\epsilon_\infty$, etc.)
were nonzero and differed greatly from asymptotic entropic quantities ($D_1$ and $H_1$) and from each other~\cite{RennerChat}.
I will illustrate with the ``second laws'' for coherence and with quasiclassical work. 

A coherence property of a quantum state $R$, called the \emph{free coherence}, has been quantified with the R\'enyi divergences $D_\alpha$~\cite{LostaglioJR14}. A thermal operation can map $R$ to $S$ only if the $D_\alpha$ decrease monotonically for all $\alpha \geq 0$. 
In the asymptotic limit, the average free coherence per particle vanishes. Schematically, 
$D_\alpha / n \to 0$~\cite{LostaglioJR14}.
Observing coherence restrictions on thermodynamic evolutions---observing a TRT prediction---involves observing nontrivial R\'enyi divergences. 

Second, the approximate work cost $\Wform(R)$ of a quasiclassical state
$R =  (\rho, H)$ depends on 
$D^\epsilon_\infty(\rho  ||  e^{ - \beta H } / Z )$, and the approximate work yield $\Wext(R)$ depends on 
$D^\epsilon_0(\rho  ||  e^{ - \beta H } / Z)$. 
In conventional statistical mechanics, the work cost and the work yield depend on the same entropic quantity. Physically realizing one-shot statistical mechanics involves observing discrepancies between R\'enyi divergences. 

We might increase our chances of observing such a discrepancy by processing a quantum state $\rho$ whose R\'enyi entropies differ greatly. 
Many-body entangled systems can occupy such states~\cite{RennerChat}. Example states include the generalization
\begin{align}   \label{eq:WState}
   \ket{\psi^W_n }
   =   \frac{1}{ \sqrt{n} } (
      \ket{1  \underbrace{ 0 \ldots 0 }_{n - 1} }  
      +  \ket{ 0 1 0 \ldots 0 }  +   \ldots   +   \ket{0 \ldots 0 1} )
\end{align}
of the W state~\cite{RennerChat}. Whereas 
$H^\epsilon_\infty( \ketbra{\psi^W_n }{\psi^W_n } )  =  0$, 
$H_0^\epsilon( \ketbra{\psi^W_n }{\psi^W_n } )  =  n$. If $n$ is large, these entropies differ greatly. Not only can one-shot statistical mechanics describe many-body systems, but entangled many-body states might facilitate tests of one-shot theory.

Conversely, one-shot theory might offer insights into entangled many-body states. Such states have been called ``exotic,'' and engineering them poses challenges. Suppose that thermal operations modeled the operations performable in a condensed-matter or optics lab. One-shot statistical mechanics would imply which states an experimentalist could transform into which, how much energy must be drawn from a battery to create a state, which coherences could transform into which, etc.

Not all one-shot results could be realized with large systems. 
A few results depend on the dimensionality of the Hilbert space on which a quantum state is defined. An example appears in~\cite[App. G.4]{Ng14}.\footnote{
The authors discuss \emph{catalysis}, the use of an ancilla to facilitate a transformation. Let $R = (\rho, H_R)$ denote a state that cannot transform into $S = (\sigma, H_S)$ by thermal operations: $R \not\mapsto S$. 
Some \emph{catalyst} $C = (\xi, H_C)$ might satisfy 
$(\rho \otimes \xi,  H_R  +  H_C)  \mapsto  (\sigma \otimes \xi,  H_S  +  H_C)$. 
Catalysts act like engines used to extract work from a pair of heat baths. 
Engines degrade, so a realistic transformation might yield $\sigma \otimes \tilde{\xi}$, wherein $\tilde{\xi}$ resembles $\xi$. For certain definitions of ``resembles,'' the agent can extract arbitrary amounts of work by negligibly degrading $C$. 
Brand\~{a}o \emph{et al.} quantify this extraction 
in terms of the dimension ${\rm dim}( \mathcal{H}_C )$ 
of the Hilbert space $\mathcal{H}_C$ on which $\xi$ is defined. 
The more particles the catalyst contains, the greater the
${\rm dim}( \mathcal{H}_C )$. Such one-shot results depend on the number of particles in the system represented by ``one copy'' of $C$.}


%
%
%
%
\subsubsection{How can we test predictions of maximal efficiencies?}
\label{section:Optima}

Many TRT theorems concern the maximal efficiency with which any thermal operation can implement some transformation $R \mapsto S$. 
``Maximal efficiency'' can mean
``the most work extractable from the transformation,'' for example.
Testing maximal efficiencies poses two challenges:
(i) No matter how efficiently one implements a transformation, one cannot know whether a more efficient protocol exists. 
(ii) Optimal thermodynamic protocols tend to last for infinitely long times and to involve infinitely large baths. 
These challenges plague not only TRTs, but also conventional thermodynamics.
We can approach the challenges in at least three ways:
(A) Deprioritize experimental tests. 
(B) Calculate corrections to the predictions. 
(C) Check whether decreasing speeds and increasing baths' sizes inches realized efficiencies toward predictions.

Proving that one has implemented a transformation maximally efficiently would amount to proving a negative. Experiments cannot prove such negatives.
Imagine an agent who wishes to create some quasiclassical state $R$ from a battery and thermal operations. Suppose that creating any state $\tilde{R}$ that lies within a distance $\epsilon$ of $R$ would satisfy the agent. The least amount $\Wform(R)$ of work required to create any such $\tilde{R}$ was calculated in~\cite{FundLimits2}.
In what fashion could an experimentalist test this prediction, if able to perform arbitrary thermal operations?

The form of a thermal operation that generates a $\tilde{R}$ is implied in~\cite{FundLimits2}.\footnote{
The functional form of $\Wform(R)$ is derived in~\cite[Suppl. Note 4]{FundLimits2}.
The proof relies on Theorem 2. 
The proof of Theorem 2 specifies how a TRT agent can perform an arbitrary free unitary. Hence the thermal operation that generates a $\tilde{R}$ is described indirectly.}
One could articulate the operation's form explicitly and could perform the operation in each of many trials. 
If $\epsilon$-closeness is defined in terms of the trace distance, one could measure each trial's output, then calculate the distance between the created state and $R$. 
One could measure the work $W$ invested in each trial and could check whether 
$W = \Wform(R)$. 
Separately, one could create $\tilde{R}$'s from another thermal operation, could measure the work $W$ invested in each trial, and could check whether each $W > \Wform(R)$. Finally, one could design a thermal operation $\mathcal{E}$ that outputs a $\tilde{R}$ if sufficient work is invested, could invest $W < \Wform(R)$ in a realization of $\mathcal{E}$, and could check that the resulting state differs from $R$ by more than $\epsilon$. Yet one would not know whether, by investing $W < \Wform(R)$ in another protocol, one could create a $\tilde{R}$.

Testing optima poses problems also because optimal thermodynamic protocols tend to proceed quasistatically. Quasistatic, or infinitely slow, processes keep a system in equilibrium. Quick processes tend to eject a system from equilibrium, dissipating extra heat~\cite{YungerHalpernGDV14}. This heat, by the First Law, comes from work drained from the battery~\cite{Jarzynski97PRE}. 
Hence TRT predictions about optimal efficiencies are likely predictions of the efficiencies of quasistatic protocols. Quasistatic protocols last for infinitely long times. Long though graduate students labor in labs, their experiments do not last forever. 

Baths' sizes impede experimental tests as time does. In~\cite[Suppl. Note 1]{FundLimits2}, ``the energy of the heat bath (and other relevant quantities such as [the] size of degeneracies) [tends] to infinity.'' 
Real baths have only finite energies and degeneracies. Granted, baths are assumed to be infinitely large (and unrealistically Markovian) in statistical mechanics outside of TRTs. As such predictions must be reconciled with real baths' finiteness, so must TRT predictions.

The reconciliation can follow one of at least three paths. First, resource theorists can shrug. Shruggers would have responded to the question at my seminar with ``No, my arch-nemesis (new friend) might as well not have heard of my results, for all they'll impact his lab.'' The $\Wform$ prediction appears in the paper ``Fundamental limitations for quantum and nanoscale thermodynamics.'' 
Fundamental limitations rarely impact experiments more than the imperfectness of experimentalists' control does. 
For example, quantum noise fundamentally impedes optical amplifiers~\cite{Caves82}. 
Johnson noise, caused by thermal fluctuations, impedes amplifiers in practice, not fundamentally.
That is, cooling an amplifier can eliminate Johnson noise. 
Yet Johnson noise often outshouts quantum noise. 
As we cannot always observe fundamental, quantum noise, 
we should not necessarily expect to observe the fundamental limitations predicted with TRTs. 

Nor, one might continue, need we try to observe fundamental limitations. Fundamental limitations bound the efficiencies of physical processes. Ideal bounds on achievable quantities have been considered physics. Thermodynamics counts as physics, though some thermodynamic transformations are quasistatic. 
TRTs need no experimental tests; ignore the rest of this paper.

Yet thermodynamics has withstood tests~\cite{Joule1845}. ``Quantum-limited'' amplifiers uninhibited by Johnson noise have been constructed. Fundamental limitations have impacted experiments, so fundamental limitations derived from TRTs merit testing. 
Furthermore, testable predictions distinguish physics from philosophy and mathematics.
If thermodynamic resource theorists are doing physics, observing TRT predictions would be fitting. 
Finally, resource theorists motivate TRTs partially with experiments. 
Experimentalists, according to one argument, can control single molecules and can measure minuscule amounts of energy. 
Conventional statistical mechanics models such experiments poorly. 
Understanding these experiments requires tools such as TRTs.
Do such arguments not merit testing?
If experimentalists observe the extremes predicted with TRTs, then the justifications for, and the timeliness of, TRT research will grow.

Tests of TRT optima can be facilitated by the calculation of corrections and by experimental approaches to ideal conditions.
To compute corrections, one could follow Reeb and Wolf~\cite{ReebW14}. They derive corrections, attributable to baths' finiteness, to Landauer's Principle. [Landauer proposed that erasing one bit of information quasistatically costs an amount 
$W_{\rm L}  = k_{\rm B} T \ln (2)$ of work.] 
The authors' use of quantum-information tools could suit the resource-theory framework.
One-shot tools are applied to finite-size baths in~\cite{TajimaH14}, 
and the resource-theory framework is applied in~\cite{Woods15}.

Instead of correcting idealizations theoretically, one might approach idealizations experimentally. One can perform successive trials increasingly slowly, and one can use larger and larger baths. As speeds drop and baths grow, do efficiencies approach the theoretical bound? 
Koski \emph{et al.} posed this question about speed when testing Landauer's Principle.  Figure 2 in~\cite{Koski14} illustrates how slowing a protocol drops the erasure's cost toward $W_{\rm L}$. Approaches to the quasistatic limit and to the infinitely-large-bath limit might be used to test TRTs.

%
%
%
%
\subsubsection{Which operations should be performed to test TRT results?}
\label{section:Existence}

To test a theorem about a transformation, one should know how to implement the transformation and what to measure. Consider a prediction of the amount of work required to transform a state $R$ into a state $S$. Some thermal operation $\mathcal{E}$ satisfies $\mathcal{E}(R)  =  S$. Checking the prediction requires knowledge of the operation's form. Which Gibbs state must one create, which unitary must one implement, and which subsystem must one discard? 
What must be measured in how many trials, and how should the measurement outcomes be processed?

Proofs of TRT theorems detail thermal operations to different extents. 
Less-detailed proofs include that of Theorem 5 in~\cite{YungerHalpernR14}. 
The authors specify how a channel transforms a state 
by specifying the form of the channel's output. 
Which Gibbs state, unitary, and tracing-out the channel involves are not specified. 
Farther up the explicitness spectrum lies the proof that, at most, an amount $\Wext(R)$ of work can be extracted with accuracy $1 - \epsilon$ from a quasiclassical state $R$~\cite[Suppl. Note 4]{FundLimits2}.
The proof specifies that ``we have to map strings of [weight $1 - \epsilon$] to a subspace of our energy block.'' The agent's actions are described, albeit abstractly.
The same proof specifies another operation in greater detail, in terms of unitaries: ``For each fixed [energy] $E_S$ [of the system of interest,] we apply a random unitary to the heat bath, and identity to [the system of interest]''~\cite[Suppl. Note 2]{FundLimits2}. 

One could describe an operation's form more explicitly. For instance, one could specify how strong a field of which type should be imposed on which region of a system for how long. Such details might depend on one's platform---on whether one is manipulating a spin chain, ion traps, quantum dots, etc. Such explicitness would detract from the operation's generality. Generality empowers TRTs. Specifying a Gibbs state, a unitary, and a tracing-out would balance generality with the facilitation of physical realizations.

In particular need of unpacking are the clock, catalyst, coupling, and time-evolution formalisms. Resource theorists developed each formalism, in a series of papers, to model some physical phenomenon. Later papers, borrowing these formalisms, reference the earlier papers implicitly or briefly. The scantiness of these references expedites theoretical progress but can mislead those hoping to test the later results. I will overview, and provide references about, each formalism.

The clock formalism is detailed in~\cite{BrandaoHORS13,FundLimits2,Malabarba15}.
Many TRT theorems concern transformations $(\rho, H)  \mapsto  (\sigma, H)$ that preserve the Hamiltonian $H$. Theorems are restricted to constant Hamiltonians ``without loss of generality.''
Yet Hamiltonians change in real-life protocols, as when fields $\mathbf{B}(t)$ are quenched. Resource theorists reconcile the changing of real-life Hamiltonians with the frozenness of TRT Hamiltonians by supposing that a clock couples to the system of interest. When the clock occupies the state $\ket{ i }$, the Hamiltonian $H( t_i )$ governs the system of interest.
The composite system's Hamiltonian remains constant, so TRT theorems describe the composite. TRT theorems restricted to constant $H$'s owe their generality to the clock formalism.

Like the clock formalism, catalysis requires detailing.
Laboratory equipment such as clocks facilitates experiments without changing much. These items serve as \emph{catalysts}, according to the resource-theory framework. If TRT predictions are tested in some lab, the catalysts in the lab must be identified. Predictions should be calculated from catalytic thermal operations~\cite{BrandaoHNOW14,Ng14}. If predictions are calculated from thermal operations, the neglect of the catalysts must be justified.

\emph{Prima facie}, couplings do not manifest in TRTs. Free unitaries couple subsystems 
$R = (\rho, H)$ and $G  =  ( e^{ - \beta H_b } / Z,   H_b)$. 
No interaction term  $H_{\rm int}$ appears in $H_{\rm tot}  =  H  +  H_b$. 
But interaction Hamiltonians couple subsystems in condensed matter and quantum optics. Since condensed matter and quantum optics might provide testbeds for TRTs, the coupling formalisms must be reconciled. An equivalence between the formalisms is discussed in~\cite[Sec. VIII]{BrandaoHORS13}.

\emph{Prima facie}, TRT systems do not evolve in time. Many authors define states 
$R = (\rho, H)$ but mention no time evolution of $\rho$ by 
$U(t)  =  \exp (- \frac{i}{\hbar} H t )$. These authors focus on quasiclassical states, whose density operators commute with their Hamiltonians: 
\mbox{$[\rho, H]  =  0$.} This $H$ generates $U(t)$. 
Hence if $[\rho, U(t)]  =  0$, the states remain constant in the absence of interactions, and time evolution can be ignored. Quantum states $\rho$ that have coherences relative to the $H$ eigenbasis evolve nontrivially under their Hamiltonians. This evolution commutes with thermal operations~\cite{LostaglioJR14,LostaglioKJR15}. 
Though they might appear not to, time evolution and couplings familiar from condensed matter and optics manifest in TRTs.


In addition to the specifying the steps in thermal operations, resource theorists could specify what needs measuring, how precisely, in how many trials, 
and how to combine the measurements' outcomes, to test TRT predictions. 
For example, many TRT theorems involve \emph{$\epsilon$-approximation}.
$\epsilon$-approximation is often defined in terms of the trace distance 
$\mathcal{D}_{\rm tr}$
between quantum states. 
$\mathcal{D}_{\rm tr}$ has an operational interpretation. 
How to use that interpretation, to check the distance from an experimentally created state to the desired state, is discussed in Sec.~\ref{section:Embezzle}.
Instead of invoking the trace distance's operational interpretation,
one could perform quantum state tomography.
The preciseness with which measurements can be performed must be weighed against the error tolerance $\epsilon$ in the theorems being tested.
As another example, some TRT predictions cannot be tested, as discussed in  
Sec.~\ref{section:Optima}.
We must distinguish which predictions are testable; then specify which measurements, implemented how precisely, and combined in which ways, would support or falsify those predictions.

%
%
%
%
\subsection{Enhancing TRTs' realism}

Six adjustments could improve how faithfully TRTs model reality.
First, we could narrow the gap between 
the operations easily performable by TRT agents
and the operations easily performable by thermodynamic experimentalists.
Second, we could incorporate violations of energy conservation into TRTs:
TRTs model closed systems, whereas most real systems are open.
Section~\ref{section:Measurement} concerns measurements;
and Sec.~\ref{section:Embezzle} concerns
thermal embezzlement, a seeming violation of the First Law of Thermodynamics in TRTs.
Faulty operations feature in Sec.~\ref{section:Error}.
Finally, modeling continuous spectra would improve TRT models of classical systems and quantum environments.

%
%
%
%
\subsubsection{Should free operations reflect experimentalists' capabilities better? If so, how?}
\label{section:EasyOps}

Should resource theorists care about how easily experimentalists can implement thermal operations? Laboratory techniques advance. Even if thermal operations challenge experimentalists now, they might not in three years. Theory, some claim, waits for no experiment. Quantum error correction (QEC) illustrates this perspective. Proposals for correcting a quantum computer's slip-ups involve elaborate architectures, many measurements, and precise control~\cite{Gottesman09}. QEC theory seemed unrealizable when it first flowered. Yet the foundations of a surface code
have been implemented~\cite{Corcoles15}, and multiple labs aim to surpass these foundations. Perhaps theory should motivate experimentalists to expedite difficult operations.

On the other hand, resource theories are constructed partially to reflect limitations on labs. The agents in resource theories reflect the operationalism inspired partially by experiments. 
Furthermore, TRTs were constructed to shed light on thermodynamics. Thermodynamics evolved largely to improve steam engines: thermodynamics evolved from practicalities. By ignoring practical limitations, thermodynamic resource theorists forsake one of their goals.\footnote{
Other goals include the mathematical isolation, quantification, and characterization of single physical quantities. Want to learn how entanglement empowers you to create more states and to perform more operations than accessible with only separable states? Use a resource theory for entanglement. Want to learn how accessing information empowers you? Use the resource theory for information~\cite{HHOShort,GourMNSYH13}.}
Finally, increasing experimental control over nanoscale systems has motivated thermodynamic resource theorists. Experimentalists, an argument goes, are probing limits of physics. They are approaching regimes near the fundamental limitations described by TRTs. This argument merits testing. Physical realizations of TRTs would strengthen the justifications for developing TRTs.

Thermodynamic experimentalists disagree with resource theorists in three ways about which operations are easy: experimentalists cannot necessarily easily implement unitaries that commute with $H_{\rm tot}$. 
Nor can experimentalists create arbitrary Hamiltonians. 
Finally, experimentalists do not necessarily value work as TRT agents do. 
Redefining thermal operations could remedy these discrepancies.

TRTs agents can perform any unitary $U$ that commutes with the total Hamiltonian $H_{\rm tot}  =  H + H_{\rm b}$ of a system-and-bath composite [Eq.~\eqref{eq:UComm}]. 
Thermodynamic experimentalists cannot necessarily. 
Consider, for example, condensed matter or quantum optics. Controlling long-range interactions and generating many-body interactions can be difficult.
Two-body interactions are combined into many-body interactions.
Hence many-body interactions are of high order in two-body coupling constants.
Particles resist dancing to the tune of such weak interactions.
Coaxing the particles into dancing challenges experimentalists but not TRT agents.


In addition to implementing energy-conserving unitaries, a TRT agent can create equilibrium states $(e^{-\beta H_{\rm b}} / Z, H_{\rm b})$ relative to the heat bath's inverse temperature $\beta$ and relative to any Hamiltonian $H_{\rm b}$ (though $H_{\rm b}$ is often assumed to have a bounded, discrete spectrum).
Experimentalists cannot create systems governed by arbitrary $H_{\rm b}$'s.
Doing so would amount to fabricating arbitrary physical systems. 
Though many-body and metamaterials experimentalists engineer exotic states, they cannot engineer everything. 

Most relevantly, experimentalists cannot construct infinitely large baths. 
Let ${\rm dim}( \mathcal{H}_b )$ denote the dimension of the Hilbert space $\mathcal{H}_b$
on which the bath's quantum state is defined.
According to the often-used prescription in~\cite{FundLimits2}, 
the output of a thermal channel approaches the desired state $R = (\rho, H)$ in the limit as   
${\rm dim}( \mathcal{H}_b ) \to \infty$. 
Real baths' states are not defined on infinitely large Hilbert spaces.
Baths' finiteness merit incorporation into TRTs.

Just as experimentalists cannot construct arbitrary bath Hamiltonians $H_b$, 
experimentalists cannot construct arbitrary ancillary Hamiltonians.
Work, for example, is often defined in TRTs with a two-level work bit~\cite{FundLimits2}.
Suppose that an agent wishes to transform $R$ into $S$ by investing an amount $W$ of work.
The agent borrows a work bit whose gap equals $W$:
$B_W  =  ( \ketbra{W}{W},   W \ketbra{W}{W} )$.
An experimentalist should prepare $B_W$ to mimic the agent. 
But an experimentalist cannot necessarily tune
a two-level system's gap to $W$.
Resource-theory predictions might require recalculating in terms of 
experimentally realizable ancillas.

Whereas TRT agents can easily implement unitaries and create states that experimentalists cannot, some thermodynamic experimentalists can easily spend work that TRT agents cannot. For example, consider a TRT agent who increases the strength $B(t)$ of a magnetic field $\mathbf{B}(t)$ imposed on a spin system governed by a Hamiltonian $H(t)$. The Hamiltonian's evolution is modeled with the clock formalism in~\cite{BrandaoHORS13,FundLimits2}. From this formalism, one can calculate the minimum amount $W_{\rm cost}$ of work required to increase $B(t)$. 

The experimentalist modeled by the agent expends work to increase $B(t)$. One can strengthen a magnetic field by strengthening the current that flows through a wire. One strengthens a current by heightening the voltage drop between the wire's ends, which involves strengthening the electric field at one end of the wire, which involves bringing charges to that end. The charges present repel the new charges, and overcoming the repulsion requires an amount $W_{\rm cost}'$ of work. Yet thermodynamic experimentalists do not necessarily cringe at the work cost of strengthening a field. 
They take less pride in charging a battery than in engineering many-body and long-range interactions, in turning a field on and off quickly, and in sculpting a field's spatial profile $\mathbf{B}(\mathbf{r})$. 
The work cast by resource theorists as valuable has less value, in some thermodynamic labs, than other resources.

Regardless of whether they value work, experimentalists might be able to measure $W_{\rm cost}'$~\cite{Pekola15}. This cost can be compared with the predicted value $W_{\rm cost}$. TRT results might be tested even if the theory misrepresents some priorities of some thermodynamic experimentalists.\footnote{
Such experimentalists value coherence similarly to TRT agents: in experiments, coherent entangled states offer access to never-before-probed physics. In quantum computers, entanglement speeds up calculations. TRT agents value coherence because they can catalyze transformations with coherent states~\cite{Aberg14}. Agents can also ``unlock'' work from coherence~\cite{KorzekwaLOJ15}. }

Resource theorists might incorporate experimentalists' priorities and challenges into TRTs. The set of free unitaries, and the set of Hamiltonians $H_{\rm b}$ relative to which agents can create Gibbs states, might be restricted~\cite{JenningsEmail}. 
One might calculate the cost of implementing a many-body interaction. 
Corrections might be introduced into existing calculations.
Additionally, alternative battery models might replace work bits.

Precedents for altering the definition of free operations exists: thermal operations are expanded to \emph{catalytic thermal operations} in~\cite{BrandaoHNOW14,Ng14}. Experimentalists use engines, clocks, and other equipment that facilitates transformations without altering (or while altering negligibly). These catalysts, Brand\~{a}o \emph{et al.} argue, merit incorporation into free operations. 
Batteries are similarly incorporated into free operations in~\cite{SkrzypczykSP13Extract}. 
Inspired by~\cite{SkrzypczykSP13Extract}, {\AA}berg redefined free operations to expose how coherences catalyze transformations~\cite{Aberg14}. 

Limitations on experimentalists' control have begun infiltrating TRTs. Brand\~{a}o \emph{et al.} showed that each of many thermal operations $\mathcal{E}_1, \mathcal{E}_2, \ldots$ can implement $R \mapsto S$~\cite{BrandaoHORS13}. If an agent can implement any $\mathcal{E}_i$, the agent can perform the transformation. Performing a particular $\mathcal{E}_i$---controlling particular aspects of the operations---is unnecessary~\cite[Suppl. Note 7]{BrandaoHORS13}.
Wilming \emph{et al.} compared TRT agents with agents who can only thermalize states entirely (can only replace states with Gibbs states)~\cite{WilmingGE14}. Real experimentalists lie between these extremes: they can perform some, but not all, thermal operations apart from complete thermalization. Following these authors, resource theorists might improve the fidelity with which free operations reflect reality.

%
%
%
%
\subsubsection{Can nonconservation of energy model systems' openness and model the impreciseness with which unitaries are implemented?}
\label{section:Nonconservation}

TRTs describe closed systems, whose energies are conserved. 
Every free unitary $U$ commutes with the total Hamiltonian:
\begin{align}
\label{eq:Cons2}
   [U, H_{\rm tot}]  =  0.
\end{align}
Experimental systems are open and have not-necessarily-conserved energies.
We might reconcile the closedness of TRT systems with the openness of experimental systems by modifying the constraint~\eqref{eq:Cons2}. 
A modification could model the impreciseness with which unitaries can be implemented~\cite{BrandaoHORS13}.

All subsystems of our universe interact with other subsystems. 
A quantum system $\mathcal{S}$ in a laboratory might couple to experimental apparatuses, air molecules, etc.  
Batteries $\mathcal{B}$, clocks $\mathcal{C}$, and catalysts $C$ have been studied in TRTs. The energy of 
$\mathcal{S} + \mathcal{B}  +  \mathcal{C}  +  C$ can more justifiably be approximated as conserved than the energy of $\mathcal{S}$ can. 
Yet the energy of $\mathcal{S} + \mathcal{B} + \mathcal{C}  +  C$ can change. 
If light shines on the clock, photons interact with $\mathcal{C}$. 
We cannot include in our calculations all the degrees of freedom in a closed, conserved-energy system. 
To compromise, we might incorporate into TRTs the nonconservation of the energies of $\mathcal{S} + \mathcal{B} + \mathcal{C}  +  C$.

Brand\~{a}o \emph{et al.} introduced nonconservation of energy into TRTs~\cite[App. VIII]{BrandaoHORS13}.
The authors suppose that work is extracted from $\mathcal{S}$ during $n$ cycles. 
The noncommutation of $U$ with $H_{\rm tot}$, they show, corrupts the clock's state. 
This corruption disturbs the work extraction negligibly in the limit as $n \to \infty$. 
Even if unable to implement $U$ precisely, the authors conclude, an agent can extract work effectively. 
The authors' analysis merits detailing and invites extensions to noncyclic processes.

In addition to modeling lack of control, a relaxation of energy conservation would strengthen the relationship between TRTs and conventional statistical mechanics.
As {\AA}berg writes in~\cite{Aberg14}, ``it ultimately may be desirable to develop a generalization which allows for small perturbations of the perfect energy conservation, i.e., allowing evolution within an energy shell, as it often is done in statistical mechanics.''

%
%
%
%
\subsubsection{How should TRTs model measurements?}
\label{section:Measurement}

The most general process modeled by quantum information theory consists of a preparation procedure, an evolution, and a measurement~\cite{NielsenC10}. 
A general measurement is represented by a \emph{positive operator-valued measure} (POVM). A POVM is a set $\mathscr{M}  =  \{ M_i \}$ of measurement operators $M_i$. The probability that some measurement of a quantum state $\rho$ yields outcome $i$ equals Tr$(\rho M_i)$. 

Measurements are (postselection is) probabilistic, whereas thermal operations are deterministic. 
The probability that some thermal operation $\mathcal{E}$ transforms some state $R = (\rho, H_R)$ into some state $S = (\sigma, H_S)$ equals one or zero. 
Not even if supplemented with a battery can thermal operations implement probabilistic transformations. 
The absence, from TRTs, of the measurement vertebra in the spine of a realistic protocol impedes the modeling of real physical processes. 

The absence also impedes the testing of TRTs. We extract data from physical systems  by measuring them. We measure quantum systems through probabilistic transformations. If we attribute to a classical statistical mechanical system a distribution over possible microstates, a measurement of the system's microstate is probabilistic. If TRTs do not model measurements, can we test TRTs? 

Alhambra \emph{et al.} propose a compromise that I will recast~\cite{Alhambra15}.  
Suppose that an agent wishes to measure a state 
$\rho' = p \sigma + (1 - p) X$ of a system $\mathcal{S}$, assuming that $\sigma$ and $X$ denote density operators.
Suppose that the agent has borrowed a memory $\mathcal{M}$, 
initialized to a nonfree pure state $(\ketbra{0}{0}, \id_d)$, from some bank.
$\mathcal{M}$ is governed by a totally degenerate Hamiltonian $\id_d$ defined on a $d$-dimensional Hilbert space. 
A free unitary $U_{\rm record}$ can couple $\mathcal{S}$ to $\mathcal{M}$.
Measuring $\mathcal{M}$ would collapse the state of $\mathcal{S}$ onto $\sigma$ or onto $X$, implementing a quantum POVM.

To incorporate the memory measurement into TRTs, we might add paid-for measurements of quasiclassical states into the set of of allowed operations. 
The agent should be able to measure a quasiclassical memory,
then reset the memory to some fiducial state by investing work.\footnote{
One might worry that $\mathcal{M}$ might not occupy a quasiclassical state after coupling to $\mathcal{S}$. But $\mathcal{M}$ has a totally degenerate Hamiltonian $\id_d$. If $\rho_M$ has coherences relative to the energy eigenbasis, free unitaries can eliminate them~\cite{GourMNSYH13}.}
Schematically,
\begin{align}
   ( \rho_M,   \id_d)   +   ( \ketbra{W}{W},   W \ketbra{W}{W} )
   \mapsto   ( \ketbra{0}{0},   \id_d )   +   ( \ketbra{0}{0},   W \ketbra{W}{W} ).
\end{align} 
The agent originally ascribes to the memory a distribution over possible pure states. Observing which microstate $\mathcal{M}$ occupies, the agent gains information transformable into work~\cite{Szilard29}. Gaining work deterministically from free operations contradicts the spirit of the Second Law of Thermodynamics. 
But the agent forfeits this work to erase $\mathcal{M}$. 
The agent must return $\mathcal{M}$, restored to its original state, to the bank.\footnote{
One might object that the measurement could project the memory's quantum state onto a pure state. Transforming any pure state into $\ket{0}$ costs no work: 
The transforming unitary commutes with the Hamiltonian $\id_d$ and so is free~\cite{GourMNSYH13}. 
But the agent could be fined the work that 
one would need to erase $\mathcal{M}$ if one refrained from measuring $\mathcal{M}$.
Imagine that a ``measurement bank'' implements measurements:
The agent hands $\mathcal{M}$ to a teller. 
Depending on the memory's state, the teller and agent agree on a fee, 
which the agent pays with a charged battery.
The teller measures $\mathcal{M}$; announces the outcome; resets $\mathcal{M}$; 
stores the battery's work contents in a vault; and returns the reset memory $( \ketbra{0}{0}, \id_d)$ and the empty battery $(\ketbra{0}{0},   W \ketbra{W}{W})$ to the agent.}

Questions about this measurement formalism remain. First, suppose that $U_{\rm record}$ correlates $\mathcal{S}$ with $\mathcal{M}$ perfectly. The memory's reduced state is maximally mixed, having the spectrum 
$P_\mathcal{M}  =  (\frac{1}{d}   \ldots   \frac{1}{d})$. 
Landauer's Principle suggests that erasing $\mathcal{M}$ costs at least an amount
$W = \kB T  \log(d)$ of work~\cite{Landauer61}. 
Instead, suppose that $U_{\rm record}$ correlates $\mathcal{S}$ with $\mathcal{M}$ imperfectly. $P_\mathcal{M}$ might not be maximally mixed;
erasing $P_\mathcal{M}$ could cost an amount 
$W( P_\mathcal{M} )  <  \kB T  \log(d)$ of work~\cite{GourMNSYH13}. 

Second, measuring and erasing $n$ copies of $\mathcal{M}$ individually costs an amount
$n \, W( P_\mathcal{M} )$ of work. 
But the agent might prefer paying a lump sum to paying for each copy individually. 
In the limit as $n \to \infty$, 
$( P_\mathcal{M} )^{\otimes n}$ can be compressed into 
$n \, H_{\rm S}( P_\mathcal{M} ) $ bits. 
Delaying payment might save the agent work~\cite{MatteoChat}.
Delayed payments, like imperfect $U_{\rm record}$'s, as well as imperfect measurements and nondegenerate memories, merit consideration.

A precedent exists for incorporating paid-for measurements into allowed operations: 
thermal operations have been expanded to catalytic thermal operations~\cite{BrandaoHNOW14,Ng14}. Just as catalysts are not free, measurements are not: 
an agent could pay to restore a degraded catalyst to its initial state and can pay to erase a memory. As catalysis has been incorporated into allowed operations, so might measurements. 
The term \emph{$\epsilon$-deterministic thermal operations} has already appeared~\cite{SalekW15}.
$\epsilon$-determinism surfaces in a related framework, 
if not in the resource-theory framework, in~\cite{Aberg13}. 
Navascu\'{e}s and Garc\'{\i}a-Pintos treat measurement similarly to Alhambra \emph{et al.}
when studying the ``resourcefulness'' of nonthermal operations~\cite{NavascuesGP15}. 

On the other hand, physically realizing TRTs might not necessitate 
the incorporation of measurements into TRTs. 
An experimenter could test a work-extraction theorem, 
derived from the TRT framework, as follows:
First, the experimenter extracts work by implementing a thermal operation.
Then, the experimenter performs a measurement absent from the TRT framework. 
The experimenter imitates two agents: 
One, a TRT agent, extracts work by a thermal operation.
This first agent passes the resultant system  to someone who ``lives outside the resource theory,'' 
whom energy constraints do not restrict.
The second agent measures the system.
Hence one might test TRTs that do not model measurements.
But externalizing measurements renders TRTs incomplete models of simple thermodynamic processes.
To model physical reality, we must refine the TRT representation of measurements.

%
%
%
%
\subsubsection{Can a measure of $\epsilon$-closeness lead to testable predictions and away from embezzlement?}
\label{section:Embezzle}

Many TRT results concern \emph{$\epsilon$-approximations} (e.g.,~\cite{FundLimits2,BrandaoHNOW14,GourMNSYH13,YungerHalpernR14,YungerHalpern14,Ng14,Alhambra15,NavascuesGP15}): Suppose that an agent wishes to create a \emph{target state} $R = (\rho, H)$. The agent might settle for some 
$\tilde{R}  =  (\tilde{\rho}, H)$ whose density operator $\tilde{\rho}$ lies within a distance $\epsilon \in [0, 1]$ of $\rho$: 
$\mathcal{D} ( \rho,  \tilde{\rho} )   \leq  \epsilon$. 
This $\epsilon$ has been called the \emph{error tolerance}, \emph{failure probability}, and \emph{smoothing parameter}. 
The distance measure $\mathcal{D}$ is often chosen to be the trace distance 
$\mathcal{D}_{\rm tr}$. 
$\mathcal{D}_{\rm tr}$ has an operational interpretation that could facilitate experimental tests. But $\mathcal{D}_{\rm tr}$ introduces \emph{embezzlement} into TRTs. Thermal embezzlement is the extraction of arbitrary amounts of work from the negligible degradation of a catalyst. A \emph{catalyst} consists of equipment, such as an engine, that facilitates a transformation while remaining unchanged or almost unchanged. 
Negligible degradation---the cost of embezzlement---is difficult to detect.
Extracting work at a difficult-to-detect cost contradicts the spirit of the First Law of Thermodynamics. Resource theorists have called for eliminating embezzlement from TRTs by redefining thermal operations or by redefining $\epsilon$-approximation~\cite{RennerChat,Ng14,OppenheimOpen}.
Three redefinitions have been proposed.
On the other hand, embezzlement might merit physical realization.

I will begin with background about the trace distance, defined as follows~\cite{NielsenC10}.
Let $\rho$ and $\tilde{\rho}$ denote density operators defined on the same Hilbert space. The \emph{trace distance} between them is
\begin{align}
   \mathcal{D}_{\rm tr} (\rho, \tilde{\rho})  :=
   \frac{1}{2} \Tr | \rho - \tilde{\rho} |,
\end{align}
wherein $| \sigma |  :=  \sqrt{ \sigma^\dag \sigma }$ for an operator $\sigma$. If 
$\mathcal{D} (\rho, \tilde{\rho})   \leq   \epsilon$, $\rho$ and $\tilde{\rho}$ are called \emph{$\epsilon$-close}. 
States $R = (\rho, H)$ and $\tilde{R} = (\tilde{\rho}, H)$ are called \emph{$\epsilon$-close} if they have the same Hamiltonian and if their density operators are $\epsilon$-close.

The trace distance has the following operational interpretation~\cite{NielsenC10}. Consider a family of POVMs parameterized by $\gamma$:
$\mathscr{M}^\gamma  =   \{ M_i^\gamma \}$.  
Consider picking a POVM (a value of $\gamma$), preparing many copies of $\rho$, and preparing many copies of some approximation $\tilde{\rho}$. 
Imagine measuring each copy of $\rho$ with $\mathscr{M}^\gamma$ and measuring each copy of $\tilde{\rho}$ with $\mathscr{M}^\gamma$.
A percentage $\Tr( M_i^\gamma  \rho )$ of the $\rho$ trials will yield outcome $i$, as will a percentage $\Tr (  M_i^\gamma  \tilde{\rho} )$ of the $\tilde{\rho}$ trials. 
The difference  $| \Tr( M_i^\gamma  \rho )   -   \Tr (  M_i^\gamma  \tilde{\rho} ) |$
between the $\rho$ percentage and the $\tilde{\rho}$ percentage will maximize for some 
$i  =  i_{\rm max}$. 
Imagine identifying the greatest difference 
$| \Tr(  M_{i_{\rm max}}^\gamma   \rho )   -   \Tr (  M_{i_{\rm max}}^\gamma \tilde{\rho} ) |$ 
for each POVM $\mathscr{M}^\gamma$. 
The largest, across the $\gamma$-values, of these greatest differences equals the trace distance between $\rho$ and $\tilde{\rho}$:
\begin{align}   \label{eq:TraceDist}
   \mathcal{D}_{\rm tr}( \rho, \tilde{\rho} )  =
   \max_\gamma   \:   
  | \Tr(  M_{i_{\rm max}}^\gamma   \rho )   -   \Tr (  M_{i_{\rm max}}^\gamma \tilde{\rho} ) |.
\end{align}
This operational interpretation of the trace distance might facilitate tests of
TRT results about $\epsilon$-approximations defined in terms of $\mathcal{D}_{\rm tr}$.

As an example, consider checking the prediction that, from an amount $\Wform(R)$ of work, thermal operations can generate an approximation $\tilde{R} = (\tilde{\rho}, H)$ to a quasiclassical $R$~\cite{FundLimits2}. Suppose we identify a thermal operation $\mathcal{E}$ expected to transform the work into a $\tilde{R}$. Suppose we can perform $\mathcal{E}$ in a lab. 
Suppose we have identified the POVM $\mathscr{M}^\gamma$ that achieves the maximum in Eq.~\eqref{eq:TraceDist}.\footnote{
\label{footnote:Eps}
Which $\mathscr{M}^\gamma$ achieves the maximum follows from the forms of $R$ and $\tilde{R}$. The experimentalist chooses the form of $R$. 
The form of the $\tilde{R}$ produced by the thermal operation is described in~\cite[Suppl. Note 4]{FundLimits2}.
(I have assumed that the $\tilde{R}$ produced by the thermal operation is the state produced in the experiment. 
But no experiment realizes a theoretical model exactly.
This discrepancy should be incorporated into calculations of errors.)}
We can test the theorem as follows: 
implement $\mathcal{E}$ on each of many copies of the work resource. Measure, with $\mathscr{M}^\gamma$, the state produced by each $\mathcal{E}$ implementation. Note which percentage of the measurements yields outcome $i$, for each $i$. Prepare many copies of $R$. Measure each copy with $\mathscr{M}^\gamma$. Note which percentage of the measurements yields outcome $i$. Identify the $i$-value $i_{\rm max}$ for which the $R$ percentage differs most from the $\mathcal{E}$ percentage. Confirm that the difference $\Delta$ between these percentages equals, at most, $\epsilon$.\footnote{
I have ignored limitations on the experimentalist. 
For example, I have imagined that $\mathcal{E}$ can be implemented infinitely many times and that infinitely many copies of $\rho$ can be prepared. They cannot. 
These limitations should be incorporated into the error associated with $\Delta$.}

Though blessed with an operational interpretation, the trace distance introduces embezzlement into TRTs~\cite{BrandaoHNOW14,GourMNSYH13,Ng14}. Suppose that thermal operations cannot transform $R = (\rho, H_R)$ into $S = (\sigma, H_S)$: 
$R  \not\mapsto S$. $R$ might transform into $S$ catalytically. 
We call $C  =  (\xi,  H_C)$ a \emph{catalyst} if $R \not\mapsto S$ 
while some thermal operation can transform
the composition of $R$ and $C$ into the composition of $S$ and $C$: 
$(R + C)  \mapsto  (S + C)$. The catalyst resembles an engine that facilitates, but remains unchanged during, the conversion of heat into work.

Realistic engines degrade:
\begin{align}   \label{eq:EpsCat}
   (R + C)  \to  (S + \tilde{C}).
\end{align} 
Suppose that the Hamiltonians are completely degenerate: $H_R = H_S = H_C = 0$. 
For every $R$ and $S$, there exist a $C$ and an arbitrarily similar $\tilde{C}$ that satisfy~\eqref{eq:EpsCat}. 
The more work the $R$-to-$S$ conversion requires, the larger $C$ must be 
[the greater the dimension ${\rm dim}(\mathcal{H}_C)$ 
of the smallest Hilbert space $\mathcal{H}_C$ on which $\xi$ can be defined]~\cite[App. G.2]{BrandaoHNOW14}. 
The required work is extracted from $C$; this extraction is called \emph{thermal embezzlement}. 
Embezzlement degrades $C$ to $\tilde{C}$. 
But if ${\rm dim}(\mathcal{H}_C)$ is large enough, 
the final catalyst state $\tilde{\xi}$ remains within trace distance $\epsilon$ of $\xi$. 
Embezzlement also in the context of nondegenerate Hamiltonians has been studied~\cite{BrandaoHNOW14,Ng14}.

Embezzlement contradicts the spirit of the First Law of Thermodynamics:
embezzlement outputs an arbitrary amount of work at a barely detectable cost.
Hence theorists have called for the elimination of embezzlement from TRTs.
Three strategies have been proposed: 
catalysts' sizes and energies might be bounded; 
$\epsilon$ might be defined in terms of catalysts' sizes; 
or $\epsilon$ might be defined in terms of a distance measure other than $\mathcal{D}_{\rm tr}$.
On the other hand, the challenge of realizing embezzlement may appeal to experimentalists,
whose violation of the First Law's spirit would appeal to theorists.

Agents cannot borrow arbitrarily large, or arbitrarily energetic, catalysts in~\cite{Ng14}. Suppose that an agent wishes to catalyze a transformation in which all Hilbert spaces' dimensions, and all Hamiltonians' eigenvalues, are bounded. 
The agent in~\cite{Ng14} can borrow only catalysts whose Hilbert spaces are small:
${\rm dim}( \mathcal{H}_C )  \leq  d_{\rm bound}$, for some fixed value $d_{\rm bound}$. Arbitrary amounts of work cannot be embezzled from such catalysts.
Suppose that the agent wants a catalyst $C$ whose spectrum is unbounded (e.g., a harmonic oscillator). Suppose that the partition function 
$Z_C  :=  {\rm Tr}( e^{- \beta H_C } )$ associated with the catalyst's Hamiltonian is finite. 
The catalysts of this sort that the agent can borrow have bounded average energies:
$\Tr ( \xi H_C )  \leq  E_{\rm bound}$.
The agent cannot embezzle from these catalysts.
In the context of bounded-spectrum catalysts, and in the context of unbounded-spectrum catalysts associated with finite $Z$'s, free operations in~\cite{Ng14} are designed to preclude embezzlement.

These limitations on catalysts' sizes and energies have pros and cons. 
The finite-$Z_C$ assumption ``holds for all systems for which the canonical ensemble is well-defined''~\cite{Ng14}.
The assumption seems practical.
Yet ``there will be specific cases of infinite-dimensional Hamiltonians where simply bounds on average energy do not give explicit bounds on thermal embezzling error''~\cite{Ng14}.
Examples include the hydrogen atom~\cite{NellyChat}.
The restrictions on infinite-dimensional $C$'s do not eradicate embezzling from TRTs. 
Additionally, bounds on dimension are related to bounds on energy~\cite{Ng14},
so the two restrictions form a cohesive family. 
Yet one restriction that eliminated all embezzlement would be more satisfying.

One might eliminate embezzlement from TRTs instead by incorporating catalysts' sizes into the definition of $\epsilon$-approximation. 
In~\cite[App. F4]{BrandaoHNOW14}, $\tilde{C}$ is called ``$\epsilon$-close'' to $C$ if 
\begin{align}   \label{eq:CatD}
   \mathcal{D}_{\rm tr} ( \xi,   \tilde{\xi} )   \leq   
   \frac{ \epsilon }{  \log \bm{(} {\rm dim}( \mathcal{H}_C) \bm{)} }.
\end{align}
This definition reflects the unphysicality of arbitrarily large catalysts. 
Yet the definition destroys the relevance of one-shot information measures to $\epsilon$-approximate catalysis. 
Suppose that $\epsilon$-approximation is defined in terms of the trace distance. 
Whether $R$ can be $\epsilon$-catalyzed into $S$ depends on the values of R\'{e}nyi divergences $D_\alpha$. 
Now, suppose that $\epsilon$-catalysis is defined as in Ineq.~\eqref{eq:CatD}. 
Whether $R$ can be $\epsilon$-catalyzed into $S$ depends only on the relative entropy $D_1$. 
Testing one-shot statistical mechanics involves the observation of signatures of $D_{\alpha \neq 1}$. 
Defining $\epsilon$-approximate catalysis in terms of catalysts' sizes impedes the testing of one-shot theory.

Third, $\epsilon$-approximation could be defined in terms of distance measures other than the trace distance. 
Many TRT predictions hold if the definition depends on any contractive metric~\cite[Sec. 6.1]{GourMNSYH13}. 
The \emph{work distance}, for example, has an operational interpretation 
and has relevance to one-shot entropies~\cite[App. G.3]{BrandaoHNOW14}. 
Suppose that some thermal operation maps $R + C$ to
$\widetilde{S + C}$. 
Suppose that, by investing an amount $\mathcal{D}_{\rm work}$ of work, one can map 
$\widetilde{S + C}$ to $S + C$ by a catalytic thermal operation. 
$\mathcal{D}_{\rm work}$ is called the \emph{work distance} 
between $\widetilde{S + C}$ and $S + C$. 
On the plus side, 
how much work one can extract from 
$R + C   \mapsto   \widetilde{S + C}$
depends on R\'{e}nyi divergences $D_{\alpha \neq 1}$.
On the downside, $\mathcal{D}_{\rm work}$ 
lacks information-theoretic properties 
of distance measures such as $\mathcal{D}_{\rm tr}$.
In terms of which distance measure $\epsilon$-approximation should be defined 
remains undetermined.

I have discussed strategies for eliminating embezzlement from TRTs.
Embezzlement appears to merit elimination 
because it contradicts the spirit of the First Law.
The spirit, not the letter.
Embezzlement seems physically realizable, in principle.
Detecting embezzlement could push experimentalists' abilities 
to distinguish between close-together states $C$ and $\tilde{C}$.
I hope that that challenge, 
and the chance to contradict the First Law's spirit,
attracts experimentalists.

%
%
%
%
\subsubsection{Can a definition of $\epsilon$-smoothing more naturally model errors and failure?}
\label{section:Error}

The smoothing parameter (alternatively, the error tolerance or failure probability) $\epsilon$ was introduced in Sec.~\ref{section:Embezzle}. 
I detailed the operational interpretation for the $\epsilon$ defined in terms of the trace distance. 
This operational interpretation involves an optimal POVM $\mathscr{M}^\gamma$.
By measuring $\mathscr{M}^\gamma$ in many trials, one might verify whether the $S$ outputted by an experimental implementation of some thermal operation 
is $\epsilon$-close to the desired state $R$.
I mentioned shortcomings of this verification scheme, as well as the embezzlement problem.
Another problem plagues $\epsilon$:
Though the ``error tolerance'' $\epsilon$ 
is defined in one-shot information theory, 
errors seem unable to manifest in
single shots of statistical mechanical protocols.
Another definition of ``error tolerance'' might suit TRTs better.

Let us detail how $\epsilon$ quantifies error in Sec.~\ref{section:Embezzle}. 
Suppose you wish to measure,
with the optimal $\mathscr{M}^\gamma$,
each of $n$ copies of $\rho$. 
Unable to prepare $\rho$ precisely, you create and measure $\tilde{\rho}$ instead. Your measurements' outcomes differ from the ideal measurements' outcomes in a percentage $\epsilon$ of the trials, as $n \to \infty$. $\epsilon$ quantifies the error introduced into your measurement statistics by your substitution of $\tilde{\rho}$ for $\rho$. 

The association of $\epsilon$ with error probability is justified by a limit 
in which the number $n$ of trials approaches infinity. 
No error occurs in any single trial;
nor does $\epsilon$ signify the probability 
that some trial will suffer an error.
Yet $\epsilon$ is defined in one-shot information theory, 
which describes finite numbers.\footnote{
Granted, one-shot information theory describes the simultaneous processing of finite numbers of \emph{copies of a probability distribution or quantum state. }
The finite numbers referred to above are numbers of \emph{sequential trials.}
The TRT manifestation of $\epsilon$ does not contradict one-shot information theory.
Yet the former contradicts the spirit of the latter.}
This paradox suggests that another definition of $\epsilon$ might suit TRTs better~\cite{RennerChat}. 

How can $\epsilon$ manifest in single shots? 
Data compression offers an example.
Consider compressing a random variable $X$
into the smallest possible number $k$ of bits.
Suppose that $X$ has a probability $p_i  \neq  0$ of assuming the value $x_i$,
for $i = 1, 2, \ldots, d$.
One bit occupies one of two possible states. 
Hence a set of $k$ bits occupies one of $2^k$ possible states.
We associate each possible state of $X$ 
with one possible state of the bits. 
Since $X$ assumes one of $d$ possible values, the bits must occupy one of
$2^k   \geq  d$ possible states.
We can compress $X$ into, at fewest,  $k  =  \lceil \log(d) \rceil$ bits.

If we used fewer bits, then upon decompressing, we would have some probability of failing to recover the value of $X$.
Suppose that $X$ probably does not assume the values
$x_1, x_2, \ldots, x_m$.
These values' probabilities sum to some tiny number $\epsilon$:
$p_1  +  p_2  +  \ldots  +  p_m  =  \epsilon$.
We can pretend, via a protocol called \emph{smoothing}, 
that $X$ will assume none of these values~\cite{RennerThesis}.
We can compress $X$ almost faithfully into $k'  =  \lceil  \log( d - m ) \rceil$ bits.
Decompressing, we have a probability $\epsilon$ 
of failing to recover the value of $X$, 
a probability $\epsilon$ of introducing an error into our representation of $X$.
Hence the names \emph{failure probability} and \emph{error tolerance}.
Smoothing thus has a natural interpretation in one-shot information-processing tasks.

Smoothing, as I argued, seems to lack a natural interpretation in one-shot statistical mechanics. 
This lack stems partially from resource theorists' having transplanted the definition of smoothing from information theory into TRTs.
Not all transplants bloom in their new environs.
Yet smoothing appears relevant to statistical mechanics:
Smoothing amounts to ignoring improbable events.
Improbable events are ignored in statistical mechanics:
Broken eggs are assumed never to reassemble, and smoke spread throughout a room is assumed never to recollect in a fireplace.
Since smoothing suits statistical mechanics,
but the standard definition of $\epsilon$ seems unsuited to one-shot statistical mechanics, resource theorists might tailor smoothing to TRTs.

%
%
%
%
\subsubsection{How should TRTs model continuous spectra?}
\label{section:Continuous}

Most TRT predictions concern discrete energies. 
Yet many real physical systems---quantum environments and classical systems---have continuous spectra.
We may need to extend TRT predictions to continuous spectra, to model real systems with TRTs.

In many TRT arguments (e.g.,~\cite{Janzing00,HHOLong,FundLimits2,SkrzypczykSP13Extract,BrandaoHNOW14,GourMNSYH13,YungerHalpernR14,YungerHalpern14}), spectra are assumed to be discrete. Simplicity and mathematical convenience justify the assumption. 
But mathematical convenience trades off with physicality. 
Resource theorists should check rigorously the limit as the spacing between levels vanishes. 
TRT predictions have been expected to govern continuous spectra,
but limits misbehave.

The need for continuous spectra in TRTs is discussed in~\cite{SkrzypczykSP13Extract,YungerHalpern14}. A battery is modeled as a weight that stores free energy as gravitational potential energy. The gravitational-energy operator 
$H_{\rm grav} = m g x$ depends on the position operator $x$, whose spectrum is continuous. Skrypzczyk \emph{et al.} argue that, if work is extracted from a finite number $n$ of bath qubits, the spectrum of $H_{\rm grav}$ can be approximated as discrete. One can choose for the spacing between consecutive energies to be $\mathscr{E} = \varepsilon / n^2$, for some fixed energy $\varepsilon$.
This approximation leads to an error of order $1 / n$~\cite[App. G]{SkrzypczykSP13Extract}. 
In the asymptotic limit as $n \to \infty$, the spacing $\mathscr{E}$ vanishes. Whether continuous spectra should be approximated as discrete in other TRT contexts merits investigation.

Incorporating continuous spectra into TRTs 
could facilitate the modeling of classical systems and quantum environments 
with which TRTs might be realized physically.
Many TRT results concern quasiclassical systems, 
whose density operators $\rho$ commute with their Hamiltonians $H$: $[ \rho, H ]  =  0$ 
(e.g.,~\cite{Janzing00,HHOLong,FundLimits2,SkrzypczykSP13Extract,BrandaoHNOW14,GourMNSYH13,YungerHalpernR14,YungerHalpern14}). 
Such quasiclassical results might be realized more easily than quantum results:
Classical platforms, such as DNA and colloidal particles, have been used to test 
small-scale statistical mechanics such as fluctuation relations~\cite{AlemanyR10,MossaMFHR09,ManosasMFHR09,BerutPC13,JunGB14,Blickle06,Saira12,An15}. 
These classical platforms might be recycled to realize TRTs.
Classical testbeds may have continuous spectra~\cite{JarzynskiQR15}.
Hence we may need to extend quasiclassical theorems to continuous spectra, 
or to justify the coarse-graining of real systems' continuous spectra into discrete spectra,
to harness existing platforms to realize TRTs.

Like classical systems, many environments of open quantum systems have continuous spectra~\cite{Agarwal13}.
TRTs model systems coupled to environments---to heat reservoirs~\cite{BrandaoHORS13}. 
Insofar as realistic reservoirs have continuous spectra, their TRT analogs should. 
Some reservoirs are modeled with quantum field theories, discussed in Sec.~\ref{section:QFTs}, which have continuous spectra. 
For example, consider a laser mode in a cavity, coupled to the surrounding room by leaky mirrors~\cite{Agarwal13}. 
This laser is related to the Jaynes-Cummings model,
which {\AA}berg introduced into TRTs~\cite{Aberg14}. 
The electromagnetic field outside the cavity has a continuum of frequencies. 
Modeling such QFTs---required to realize TRTs with common systems like lasers---invites us to incorporate continua into TRTs. 

%
%
%
%
\subsection{More-out-of-the-way opportunities}

Physical realizations of TRTs require confrontation of the foregoing nine challenges.
The following two challenges appear less crucial.
Yet the following could lead to realizations with physical platforms
that TRTs could not model with just adjustments discussed above.
First, modeling quantum field theories with TRTs could facilitate quantum-optics and condensed-matter realizations. 
Second, TRTs might reach physical platforms via fluctuation relations.
Fluctuation relations describe small scales, as TRTs can, and have withstood experimental tests.

%
%
%
%
\subsubsection{Can TRTs model quantum field theories?}
\label{section:QFTs}

Quantum field theories (QFTs) represent thermodynamic systems that range from lasers to condensed matter to black holes. Quantum optics and condensed matter are increasingly controllable and conscripted for quantum computation. Holography is shedding the light of quantum information on black holes~\cite{Harlow14}, to which one-shot information theory has been applied~\cite{Czech14}. 
Modeling QFTs with TRTs could unlock testbeds and applications for TRTs.
Further motivation appears in~\cite{KorzekwaLOJ15}:
The authors study the ``unlocking,'' aided by an external field, of work stored in coherence. 
The field suffers a back reaction.
Approximating the field as classical neglects the back reaction.
To calculate the unlocking's cost, one must quantize the field.

I will overview steps with which we can incorporate QFTs into TRTs: 
an introduction of Fock space into TRTs, more attention to the number operator in TRTs, and more attention to unbounded spectra. 
Instead of studying general Hamiltonians and qubits, resource theorists would need to focus on quantum-optics and condensed-matter Hamiltonians. Groundwork for these expansions has been laid in~\cite{YungerHalpernR14,YungerHalpern14,Aberg14}.

Specifying a state in a TRT involves specifying a Hilbert space. 
Quantum states in QFTs are defined on Fock spaces. 
Since Fock spaces are Hilbert spaces, TRTs offer hope for modeling QFTs. 
A popular Fock-space basis consists of eigenstates $\ket{n}$ of the particle-number operator $N$. A similar operator was introduced into TRTs in~\cite{YungerHalpernR14,YungerHalpern14}. 
The number-like operators in~\cite{YungerHalpernR14,YungerHalpern14} have bounded spectra, whereas arbitrarily many particles can populate a QFT. 
Spectra were bounded in keeping with the boundedness of the energy spectra 
in many TRT proofs. 
TRT spectra are bounded ``for convenience'' or ``for simplicity.''
This phrasing suggests that proofs are believed to extend easily to unbounded spectra.
That the proofs extend merits checking, 
and how best to model unbounded spectra in TRTs merits consideration. Continuous spectra, their relevance to QFTs, and their incarnation in TRTs is discussed in Sec.~\ref{section:Continuous}.

Unbounded spectra and lasers forayed into TRTs in~\cite{SkrzypczykSP13Extract,Aberg14,Ng14}. In~\cite{Ng14}, catalysts have unbounded spectra. The expectation value of each catalyst's Hamiltonian remains below some finite value $E$. The effective cutoff reduces the problem to a finite-spectrum problem.
The cutoff is justified with the finiteness of real systems' energies.

Cutoffs feature also in~\cite{SkrzypczykSP13Extract}: a weight (such as a stone) that has gravitational potential energy models a battery. The Hamiltonian $H_{\rm grav} = mg x$ governs the weight, wherein the position operator $x$ denotes the weight's height. The position operator has an unbounded spectrum. Yet the battery's height must lie below some cutoff. Once a weight reaches a certain height above the Earth, $mgx$ approximates the stone's potential energy poorly. 
Calculating a limit as the cutoff approaches infinity might incorporate truly unbounded spectra fundamentally into TRTs. 

In~\cite{Aberg14}, {\AA}berg treats a doubly infinite ladder and a half-infinite ladder as environments. The doubly infinite ladder's energy spectrum runs from $-\infty$ to $\infty$. Though unphysical, the ladder simplifies proofs. {\AA}berg extends these proofs to the more physical half-infinite ladder, whose energy spectrum is bounded from below but not from above. 
This harmonic oscillator could pave the TRT path toward QFTs, which consist of oscillators that vibrate at various frequencies. 
One workhorse of QFT is the Jaynes-Cummings model of quantum optics. 
The model describes how matter exchanges energy with an electromagnetic field. 
{\AA}berg models a Jaynes-Cummings-like system with his framework. 
``[T]o what extent such a generalized Jaynes-Cummings interaction can be obtained, or at least approximated, within realistic systems,'' he writes, ``is left as an open question''~\cite[App. E.2]{Aberg14}.

{\AA}berg discusses also lasers, a stalwart of many labs. Laser light is often modeled with a coherent state,
\begin{align}
   \ket{\alpha}  :=  e^{ - \frac{1}{2} | \alpha |^2}  
   \sum_{l = 0}^\infty  \frac{ \alpha^l }{ \sqrt{l!} } \ket{l},
\end{align}
wherein $\{ \ket{ l } \}$ denotes the energy eigenbasis and $\ket{\alpha}$ denotes an eigenstate of the annihilation operator $a$: $a \ket{\alpha}  =  \alpha \ket{\alpha}$~\cite{Agarwal13}.
{\AA}berg studies, rather than coherent states $\ket{ \alpha }$, 
uniform superpositions 
$\ket{ \eta_{ L, l_0 } }  :=  \sum_{l = 0}^{L - 1}  \frac{1}{\sqrt{L}}  \ket{ l_0  +  l }$
of neighboring energy eigenstates.
Extending his analysis to the $\ket{\alpha}$ can shift TRTs toward modeling lasers and other real systems described by QFTs.

%
%
%
%
\subsubsection{Can fluctuation theorems bridge TRTs to experiments?}

\emph{Fluctuation theorems} are statistical mechanical predictions 
about systems arbitrarily far from equilibrium. 
Fluctuation theorems have withstood experimental tests and describe small scales. 
Can TRTs reach experiments via fluctuation theorems? 
If so, how? I will survey progress toward answers.

Fluctuation theorems describe the deviations, from equilibrium values,
of outcomes of measurements of statistical mechanical systems.
Consider a system coupled to a bath at inverse temperature $\beta$.
Suppose that a perturbation changes the system's Hamiltonian
from $H_i$ to $H_f$.
If the system consists of a gas, it might undergo compression~\cite{Crooks99}.
If the system consist of a trapped ion, a laser might induce a time-evolving field~\cite{An15}. 
Consider time-reversing the perturbation. 
Statistics that characterize the forward process can be related to 
statistics that characterize the reverse process,
to differences $\Delta F$ between free energies,
and to deviations from equilibrium statistics.
Such relations are fluctuation theorems.

Examples include Crooks' Theorem and Jarzynski's Equality. 
Let $P_\fwd(W)$ denote the probability
that some forward trial will require an amount $W$ of work
(e.g., to compress the gas).
Let $P_\rev(-W)$ denote the probability
that some reverse trial will output an amount $W$.
According to \emph{Crooks' Theorem}~\cite{Crooks99},
\begin{align}
   \frac{  P_\fwd(W)  }{  P_\rev (-W)  }
   =  e^{ \beta (W  -  \Delta F) }.
\end{align}
This $\Delta F  =  F_f  -  F_i$ denotes the difference between 
the free energy $F_f  =  - \kB T \ln (Z_f)$ of the equilibrium state $e^{ - \beta H_f } / Z_f$ 
relative to the final Hamiltonian
and the free energy $F_i$ of the equilibrium state
relative to $H_i$.
Whereas $\Delta F$ characterizes equilibrium states, 
$W$ characterizes a nonequilibrium process.
Jarzynski's Equality informs us about nonequilibrium,
difficult to describe theoretically,
via the equilibrium $\Delta F$,
easier to describe theoretically.
Conversely, from data about nonequilibrium trials,
easier to realize in practice,
we can infer the value of the equilibrium $\Delta F$,
an unrealizable ideal.
Multiplying each side of Crooks' Theorem by $P_\rev (-W)  e^{ - \beta W }$,
then integrating over $W$, yields \emph{Jarzynski's Equality}~\cite{Jarzynski97}:
\begin{align}
   \langle   e^{ - \beta W }   \rangle
   =   e^{ - \beta \Delta F }.
\end{align}

Fluctuation theorems were derived first from classical mechanics.
They have been extended to quantum systems~\cite{CampisiHT11}
and generalized with information theory (e.g.,~\cite{SagawaU10}).
Experimental tests have involved DNA~\cite{MossaMFHR09,ManosasMFHR09,AlemanyR10}, trapped colloidal particles~\cite{Blickle06}, single-electron boxes~\cite{Saira12}, trapped-ion harmonic oscillators~\cite{An15}, and other platforms.

Fluctuation theorems and TRTs describe similar (and, in some cases, the same) problems. 
First, both frameworks describe arbitrarily-far-from equilibrium processes.
Second, each framework features work, entropy, and new derivations of the Second Law.
Third, both frameworks describe small scales.
I discussed the relationship between small scales and TRTs 
in Sec.~\ref{section:OneShot}. 
The systems that obey fluctuation theorems most noticeably are small:
Deviations from equilibrium behaviors decay with system size~\cite{Schroeder00}.
Hence we can most easily detect deviations in small systems,
such as single strands of DNA.
Fourth, heat exchanges governed by Crooks' Theorem
can be modeled with TRT thermal operations~\cite{YungerHalpernGDV14}.
Since the physics described by fluctuation theorems overlaps with the physics described by TRTs, and since experimentalists have tested fluctuation theorems, fluctuation theorems might bridge TRTs to experiments.

Construction of the bridge has begun but needs expansion. 
{\AA}berg derived a Crooks-type theorem from one-shot statistical mechanics~\cite[Suppl. Note 10B]{Aberg13}. 
How to model, with TRTs, a process governed by Crooks' Theorem 
was detailed in~\cite{YungerHalpernGDV14}. 
A fluctuation relation was derived from TRT principles in~\cite{SalekW15}. 
In~\cite{YungerHalpernGDV15},
one-shot analogs of asymptotic fluctuation-theorem work quantities 
were derived. 
These first steps demonstrate the consistency between fluctuation theorems and TRTs (or one-shot statistical mechanics). 
How to wield this consistency, to link TRTs to experiments via fluctuation theorems, remains undetermined.

%
%
%
%
\section{Conclusions}

During the past few years, the literature about thermodynamic resource theories has exploded. Loads of lemmas and reams of theorems have been proven. To what extent do they describe physical reality? Now that TRTs have matured, they merit experimental probing. I have presented eleven opportunities in physically realizing TRTs. The challenges range from philosophical to practical, from speculative to expectedly straightforward. These opportunities might generalize to physical realizations of other resource theories, such as the resource theory for coherence~\cite{WinterY15}.

I concentrated mostly on gaps in resource theories, on how theorists might nudge their work toward physical realizations. Yet I hope that the discussion will appeal to experimentalists. An experimentalist opened the Q\&A of my seminar two years ago. His colleagues and thermodynamic resource theorists have an unprecedented opportunity to inform each other. Let the informing begin\ldots preferably with more cooperation and charity than during the Q\&A.

%
%
%
%
\section*{Acknowledgements}
I am grateful to Fernando Brand\~{a}o, L\'idia del Rio, Ian~Durham, Manuel~Endres, Tobias~Fritz, Alexey~Gorshkov, Christopher~Jarzynski, David~Jennings, Matteo~Lostaglio, Evgeny~Mozgunov, Varun~Narasimhachar, Nelly~Ng, John~Preskill, Renato~Renner, Dean~Rickles, Jim~Slinkman, Stephanie~Wehner, and Mischa~Woods for conversations and feedback. This research was supported by an IQIM Fellowship, NSF grant PHY-0803371, and a Virginia Gilloon Fellowship. The Institute for Quantum Information and Matter (IQIM) is an NSF Physics Frontiers Center supported by the Gordon and Betty Moore Foundation. 
Stephanie~Wehner and QuTech offered hospitality at TU Delft during the preparation of this manuscript.
Finally, I thank that seminar participant for galvanizing this exploration.

%
%
%

\bibliographystyle{h-physrev}
\bibliography{Edd-Wheeler_Refs}


\end{document}